\documentclass[twocolumn]{aastex62}

\newcommand{\REarth}{\ensuremath{R_{\oplus}}\xspace}

\newcommand{\Kepler}{\textit{Kepler}\xspace}
\newcommand{\Kdeux}{\textit{K2}\xspace}
\newcommand{\mic}{$\mu$m\xspace}

\usepackage{amsmath}
\usepackage{supertabular}
\usepackage{array}
\usepackage{multirow}
\usepackage{graphicx}
\usepackage{txfonts}
\usepackage{amssymb}	
\usepackage{xspace}
\bibpunct{(}{)}{;}{a}{}{,}

\received{2018 July 24}
\revised{2018 August 15}
\accepted{2018 August 16}
\submitjournal{AJ}

\shorttitle{\Kdeux targets observed with SPHERE/VLT}
\shortauthors{Ligi et al.}

\begin{document}

\title{\Kdeux Targets Observed with SPHERE/VLT\,: An M4-7 Dwarf Companion Resolved around EPIC\,206011496\footnote{Based on observations collected at the European Organisation for Astronomical Research in the Southern Hemisphere under ESO programs 98.C-0779(A) and 99.C-0.279(A)}\footnote{This work is based on data products produced at the SPHERE Data Center hosted at OSUG/IPAG, Grenoble.}
}

\correspondingauthor{Roxanne Ligi}
\email{roxanne.ligi@inaf.it}

\author{Roxanne Ligi}\altaffiliation{Marie Sk\l odowska-Curie/AstroFIt2 fellow}
\affiliation{INAF--Osservatorio Astronomico di Brera, Via E. Bianchi 46, I-23807, Merate, Italy}

\author{Olivier Demangeon}
\affiliation{Instituto de Astrof\'isica e Ci\^{e}ncias do Espa\c co, Universidade do Porto, CAUP, Rua das Estrelas, 4150-762 Porto, Portugal}

\author{Susana Barros}
\affiliation{Instituto de Astrof\'isica e Ci\^{e}ncias do Espa\c co, Universidade do Porto, CAUP, Rua das Estrelas, 4150-762 Porto, Portugal}

\author{Dino Mesa}
\affiliation{INCT, Universidad De Atacama, calle Copayapu 485, Copiapó, Atacama, Chile}

\author{Mariangela Bonavita}
\affiliation{INAF--Osservatorio Astronomico di Padova, Vicolo dell’Osservatorio 5, I-35122, Padova, Italy}
\affiliation{Institute for Astronomy, The University of Edinburgh, Royal Observatory, Blackford Hill, Edinburgh, EH9 3HJ, UK}

\author{Arthur Vigan}
\affiliation{Aix Marseille Univ, CNRS, LAM, Laboratoire d'Astrophysique de Marseille, Marseille, France}

\author{Mickael Bonnefoy}
\affiliation{Univ. Grenoble Alpes, CNRS, IPAG, F-38000 Grenoble, France}

\author{Raffaele Gratton}
\affiliation{INAF -- Osservatorio Astronomico di Padova, Vicolo dell Osservatorio 5, I-35122, Padova, Italy}

\author{Magali Deleuil}
\affiliation{Aix Marseille Univ, CNRS, LAM, Laboratoire d'Astrophysique de Marseille, Marseille, France}



\begin{abstract}

The quest to discover exoplanets is one of the most important missions in astrophysics, and is widely performed using the transit method, which allows for the detection of exoplanets down to the size of Mercury. However, to confirm these detections, additional vetting is mandatory. 
We selected six \Kdeux targets from campaigns $\#1$ to $\#8$ that show transit light curves corresponding to Earth-sized to Neptune-sized exoplanets. We aim to discard some scenarios that could mimic an exoplanetary transit, leading to a misinterpretation of the data. 
We performed direct imaging observations using the SPHERE/VLT instrument to probe the close environment of these stars. For five of the \Kdeux targets, we report no detection and we give the detection limits. For EPIC\,206011496, we detect a 0.38 $\pm$ 0.06 M$_{\odot}$ companion at a separation of 977.12 $\pm$ 0.73~mas (140.19 $\pm$ 0.11 au). The spectral analysis corresponds to an M4-7 star, and the analysis of the proper motion shows that it is bounded to the primary star. EPIC\,206011496 also hosts an Earth-like planetary candidate. If it transits the primary star, its radius is consistent with that of a super-Earth. However, if it transits the companion star, it falls into the mini-Neptune regime.

\end{abstract}




\section{Introduction} 
\label{sec:Introduction}

After the discovery of thousands of exoplanets, mainly thanks to the transit and radial velocity (RV) methods, we have moved from an era of detection into an era of characterization of exoplanets. But to properly characterize a planet, one needs a measurement of its radius and mass. While the \Kepler \citep{Borucky2010} mission provided a large amount of exoplanets, the extended mission \Kdeux \citep{howellK2MissionCharacterization2014} targets brighter stars, and represents the first opportunity to massively characterize both the mass and radius of Earth-sized to Neptune-sized exoplanets.

In this context, the role of high-resolution imaging (\textsc{HRI}) is dual. First, \textsc{HRI} helps confirm the planetary nature of the detected transit. Second, it allows us to significantly reduce the possible bias of the measurement of the planetary radius \citep[e.g.,][]{Leger2009, Ciardi2015}.

Concerning the first role, \textsc{HRI} is not intented to directly confirm the nature of the transit. Confirmation is achieved through the detection of the planetary signature with another, independent observation technique (typically RV). However, when independent observations cannot constrain the planetary nature of the transits, one  can perform a probabilistic validation of the transit nature \citep[e.g.,][]{diazPASTISBayesianExtrasolar2014, Moutou2014, santernePASTISBayesianExtrasolar2015}. This consists of comparing the posterior probability of all the possible scenarios for the presence of the transit given all available data. In this context, the presence of nearby contaminant stars should be closely investigated, and in the case of very shallow transits, special attention should be given to the very close vicinity of the target. Hence, \textsc{HRI} helps answer the following question\,: is there a chance-aligned eclipsing system in the angular vicinity of the target that could mimic the transit detected in the target's light curve\,?
Given that the transit of an Earth-sized to Neptune-sized planet can be mimicked by a background eclipsing system down to 10 and 7.5 mag fainter than the target star, respectively (see the detailed calculation in Appendix~\ref{app:calculContaminants}), we need \textsc{HRI} instruments capable of reaching such contrasts within $\approx$6$\arcsec$, the typical diameter of \Kepler's broad point spread function (PSF). More specifically, \textsc{HRI} helps with identifying very close contaminants missed by classical imaging.

Concerning the second role of \textsc{HRI}, assuming that we can confirm the planetary nature of the transit, and according to Equation~(\ref{eq:transitdepth}), the presence of a contaminant would still bias the measurement of the transit depth (TD) and thus the measurement of the planetary radius.
\cite{Ciardi2015} showed that ignoring the contamination can lead to an underestimation of planetary radii up to a factor 1.5, corresponding to an overestimation of the planet bulk density of a factor $\sim3$, for the \Kepler Objects of Interest. However, they claimed that with additional \textsc{HRI}, the bias in the planetary radii underestimation drops to 1.2.

We present the observations of six \Kdeux targets performed with the VLT/SPHERE instrument from 2016 to 2017. With its capacity to detect companions of $\Delta\mathrm{mag}$ below 12 down to separations of $0\farcs1$ around stars brighter than 11 mag in the $R$ band, SPHERE \citep{Beuzit2008} is one of the few instruments capable of detecting contaminants faint enough to mimic an Earth-sized to Neptune-sized transit within \Kepler's PSF.
In Section~\ref{sec:K2ample} we present the sample of our \Kdeux targets and explain the observing modes and data reduction processes in Section~\ref{sec:obs}. Section~\ref{sec:results} describes the results of our observations, which are discussed in Section~\ref{sec:discussion}. 
 
\section{Sample Selection of \Kdeux Targets} 
\label{sec:K2ample}

Our sample comprises \Kdeux targets from campaigns $\#1$ to $\#8$. 
We selected stars whose light curve exhibits a transit-like signal of depth below 100~ppm, compatible with the transit of an Earth-sized or Neptune-sized planet (see Appendix~\ref{app:calculContaminants}). To reach the sensitivity required to detect the corresponding contaminants (see Section~\ref{sec:Introduction} and Appendix~\ref{app:calculContaminants}), we had to restrict ourselves to stars brighter than 13th magnitude in the \Kepler bandpass.

\Kdeux targets presenting transit-like events were identified using both the already published lists of transiting planetary candidates (see Table~\ref{tab:transitinfo} in Appendix~\ref{app:DetailsTargets} for references) and detections made by our team. For the latter, we first used the POLAR software \citep{barrosNewPlanetaryEclipsing2016} to reduce the target-pixel files and produce high-precision light curves. These were then searched for transit-like events with two independent analyses, as described in \citet{barrosNewPlanetaryEclipsing2016} and \citet{armstrongK2VariableCatalogue2015}.
During these searches, we checked the detected transit events against a set of diagnostics that allowed us to identify false-positives\,: even/odd TD differences, out-of-transit variations, and the presence of a secondary eclipse.

We obtained a sample of 6 stars harboring the most promising Earth-sized and Neptune-sized planetary candidates whose natures were not confirmed at the time. These targets are listed in Table~\ref{tab:transitinfo} (Appendix~\ref{app:DetailsTargets}) along with the most important information regarding the transits detected in their light curves, the corresponding size of the exoplanetary candidate, and current literature (Appendix~\ref{app:DetailsTargets}).

\section{SPHERE Observations and Data Reduction}
\label{sec:obs}

The SPHERE observations of our \Kdeux targets were performed during the ESO periods P98 and P99 through the open time programs 98.C-0779(A) and 99.C-0.279(A). The data were acquired using the IRDIFS mode, in the pupil-tracking mode with the \texttt{N$\_$ALC$\_$YJH$\_$S} coronagraph (185~mas diameter). IRDIS \citep[InfraRed Dual-band Imager and Spectrograph\,;][]{Dohlen2008} was used in the dual-band imaging mode \cite[DBI\,;][]{Vigan2010} in the $H2H3$ filters ($\lambda_{H2} = 1.587$~\mic, $\lambda_{H3} = 1.667$~\mic) and IFS \citep[Integral Field Spectrograph\,;][]{Claudi2008} in the $YJ$ bands (0.95$-$1.35~\mic, $R=50$). A description of the observations is provided in Table~\ref{tab:obslog} (Appendix~\ref{app:ObsLog}).

The data were reduced at the SPHERE Data Center \cite[DC\,;][]{Delorme2017} using the SPHERE Data Reduction and Handling software \citep[DRH\,;][]{Pavlov2008}. Bad-pixel and dark-field corrections were applied during the data treatment, as well as a frame selection using the routine offered by the DC for all targets except EPIC\,206011496. Frames for which the flux in the central spot are beyond $1\sigma$ from the median flux are rejected, which allows us to keep most good images. In general, $\sim1/3$ of the frames were removed after selection. Concerning EPIC\,206011496, we detected a companion at the edge of the IFS field of view of epoch 2017 August 14 (see Section~\ref{sec:analyseEPIC206}). The companion was within the IFS field of view for only 9 frames. These were used in the analysis. We selected the bad frames in the IRDIS data using the method described in \cite{Ligi2018} for the LAM-ADI pipeline, i.e., we excluded frames presenting a flux above or below 1.5$\sigma$ of the mean flux calculated from the moving-average of the 100 frames around the considered one. We then used the \textit{Specal} routine \citep{Galicher2018} to apply different data reduction algorithms, namely the TLOCI \citep[Template-Locally Optimized Combination of Images\,;][]{Marois2014}, PCA \citep[Principal Component Analysis\,;][]{Soummer2012}, cADI \citep[Classical Angular Differential Imaging\,;][]{Marois2006} and noADI. The different algorithms differ in how they discriminate planets from speckle patterns \citep{Delorme2017}, i.e., in their description of stellar speckles, which are then subtracted to the image. In all the algorithms that we used (except noADI), the images were then rotated to a common orientation, averaged, and mean-combined. Using these different algorithms allows us to verify that hypothetical artifacts are not interpreted as planetary candidates, or inversely, that no candidate is missed.
To confirm the results, the data of EPIC\,206011496 of epoch 2017 August 14 were also reduced with the LAM-ADI \citep{Vigan2015} and the ASDI-PCA \citep{Mesa2015} pipelines. The results are similar to those obtained with the SPHERE DRH.

\section{Results}
\label{sec:results}

\begin{figure*}[ht]
\begin{center}$
\begin{array}{ccc}
\hspace*{-0.7cm}
\includegraphics[scale=0.25]{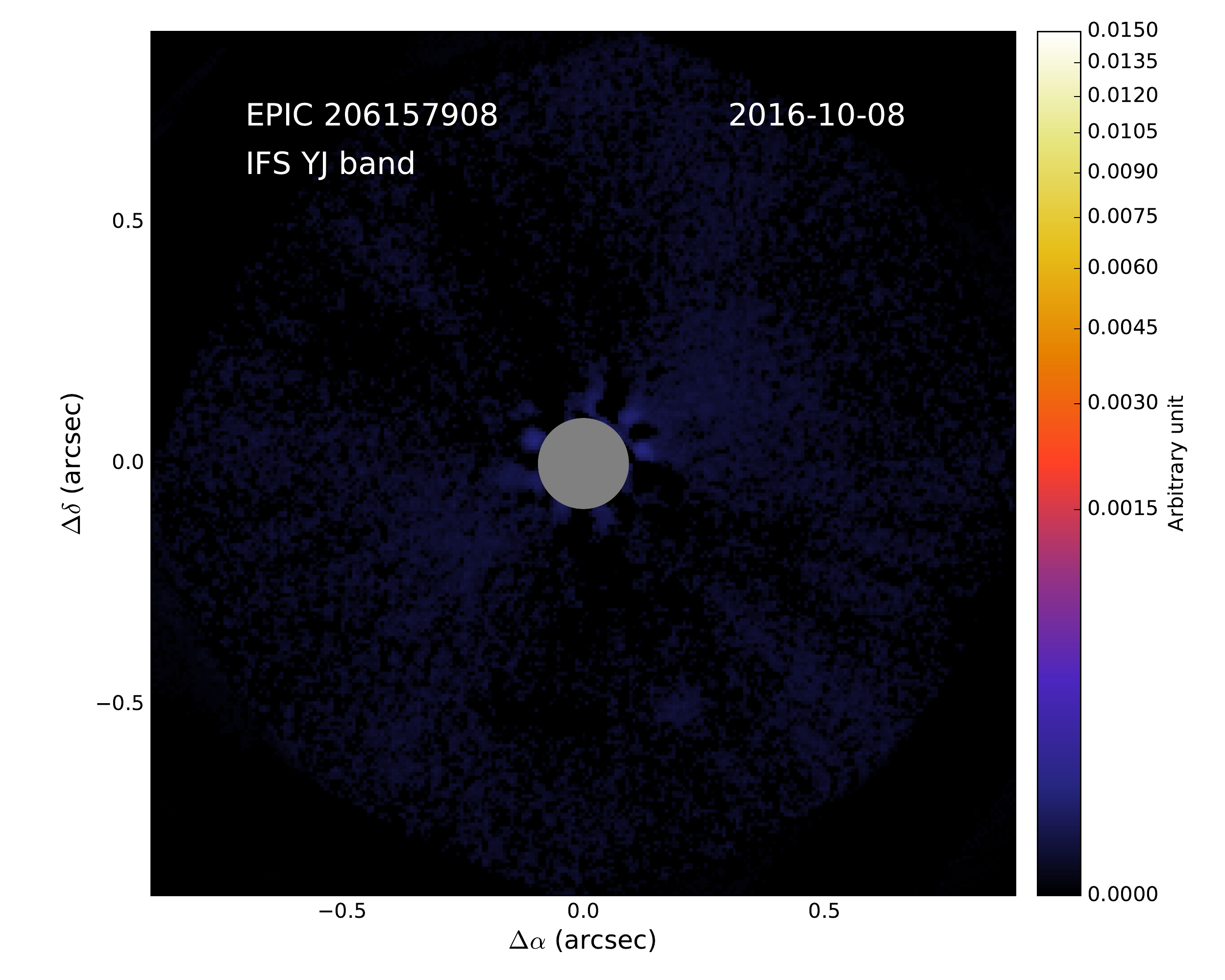} &
\hspace*{-0.5cm}
\vspace*{-0.2cm}
\includegraphics[scale=0.25]{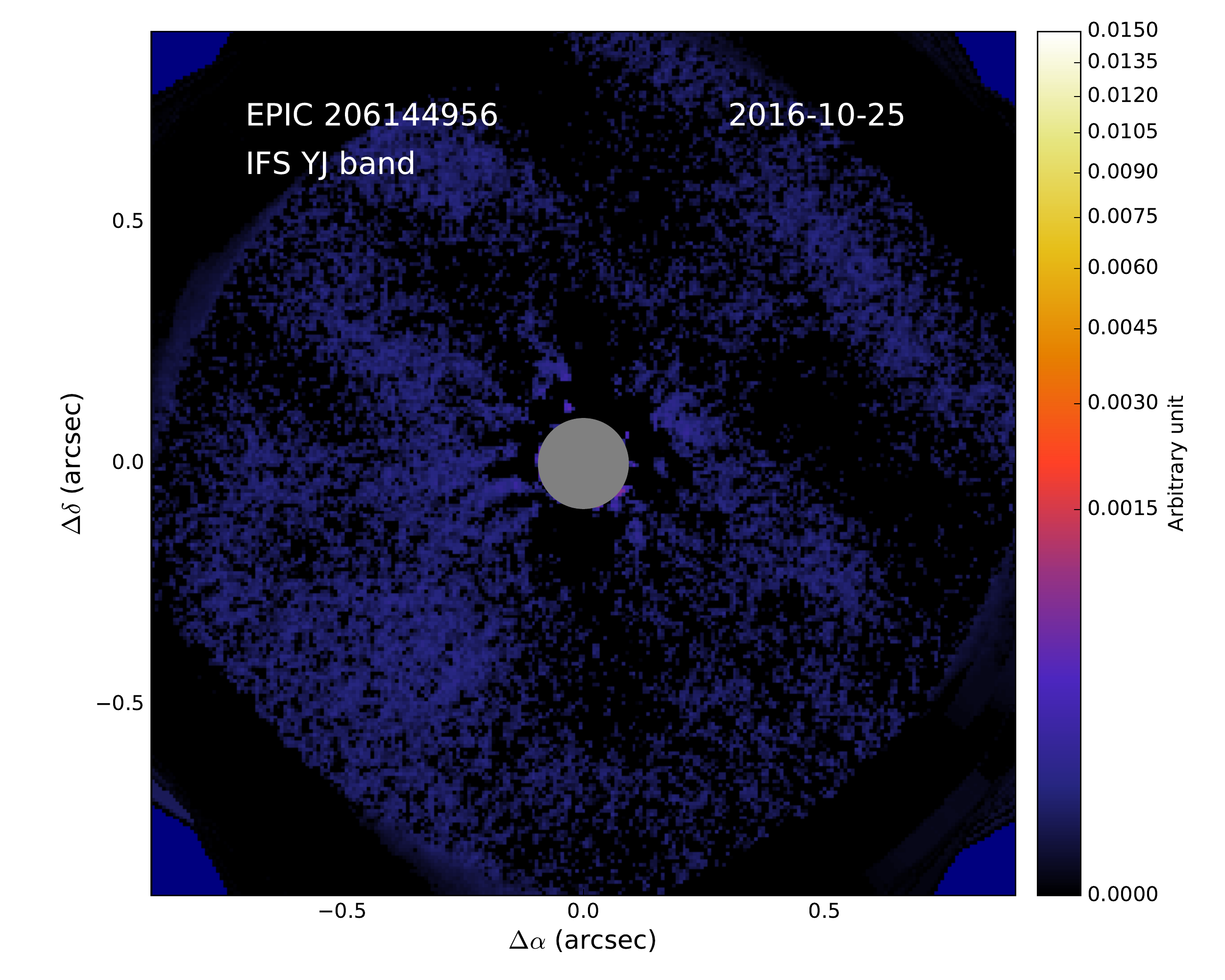} &
\hspace*{-0.5cm}
\vspace*{-0.2cm}
\includegraphics[scale=0.25]{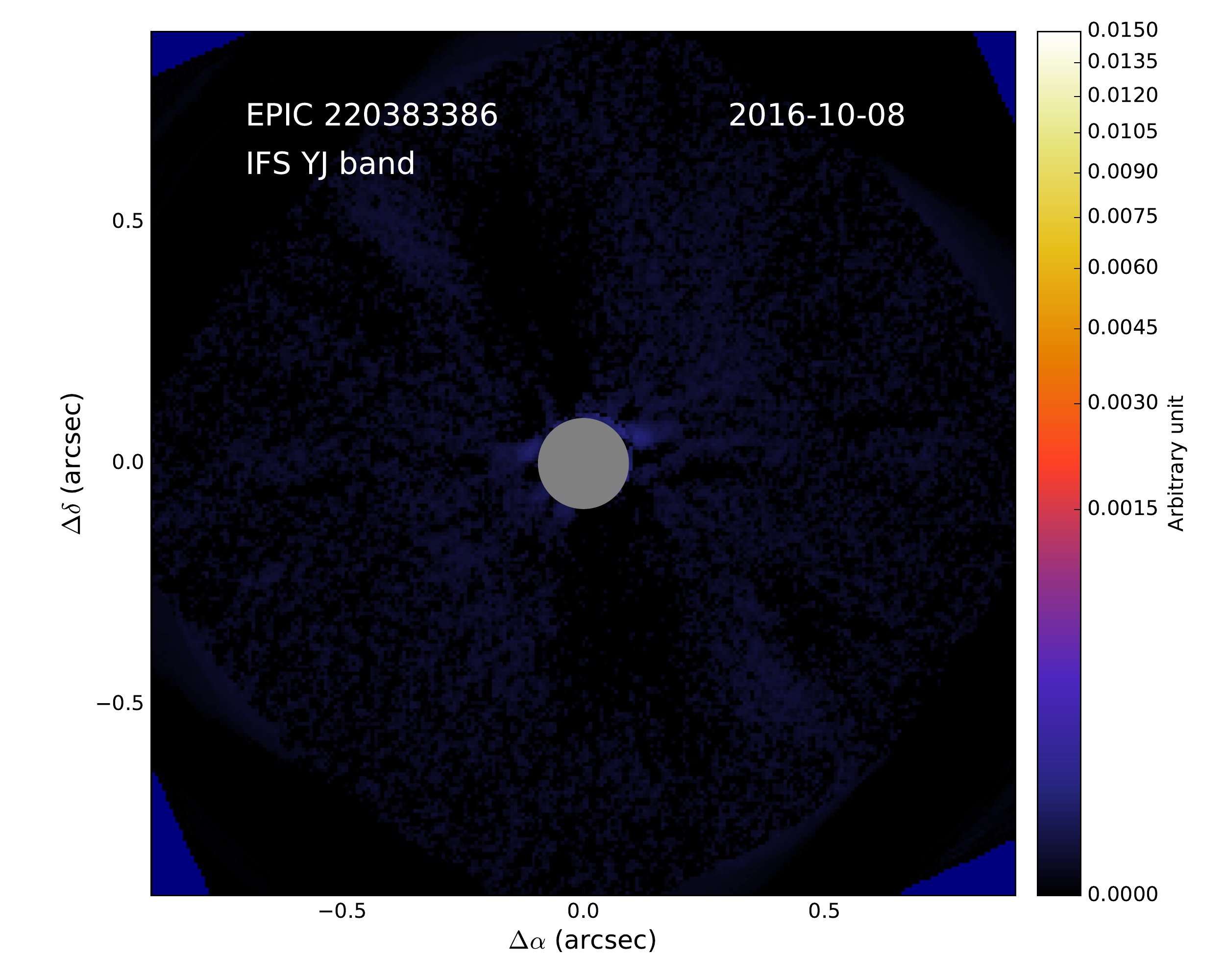} \\
\hspace*{-0.7cm}
\vspace*{-0.2cm}
\includegraphics[scale=0.25]{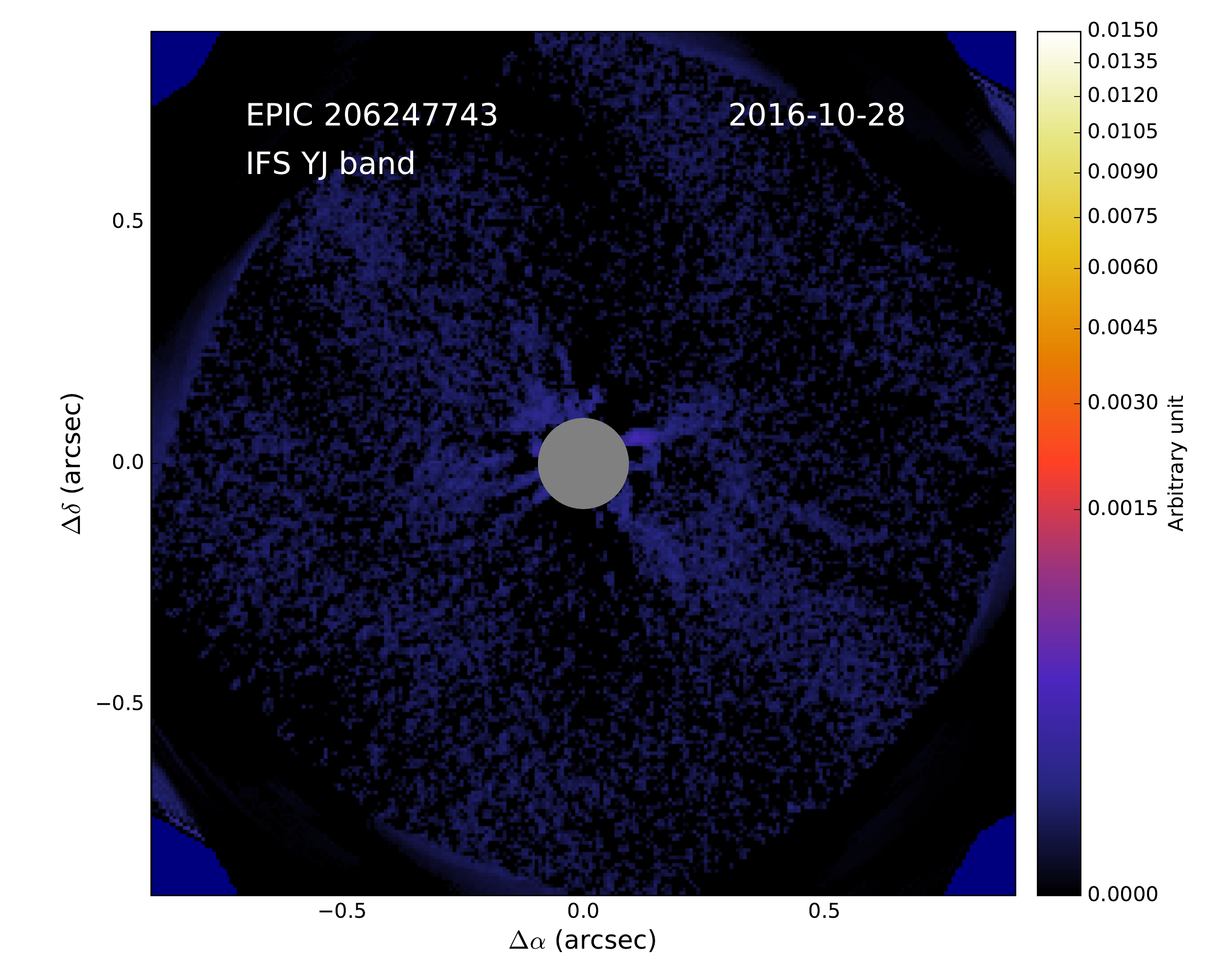} &
\hspace*{-0.5cm}
\includegraphics[scale=0.25]{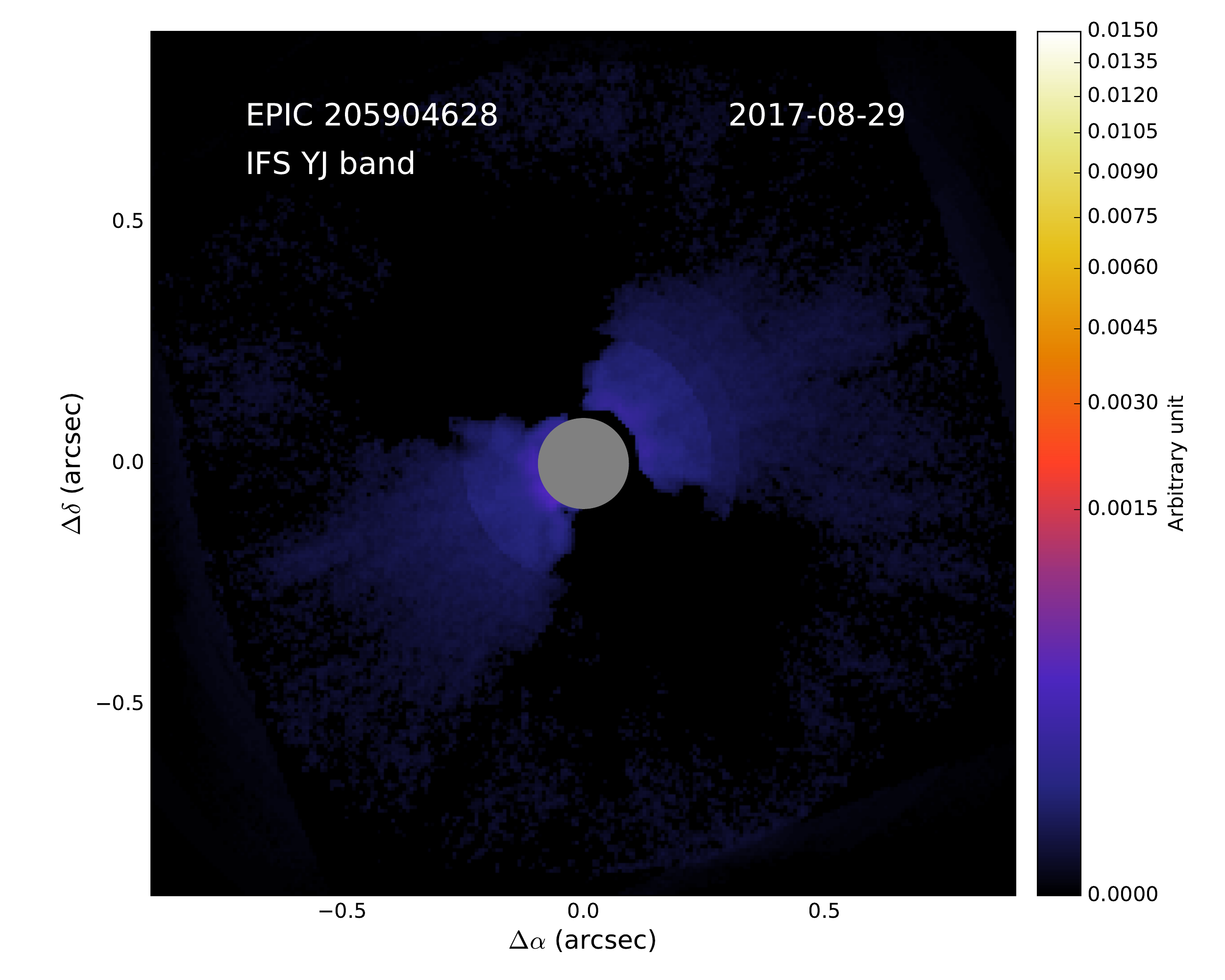} &
\hspace*{-0.5cm}
\includegraphics[scale=0.25]{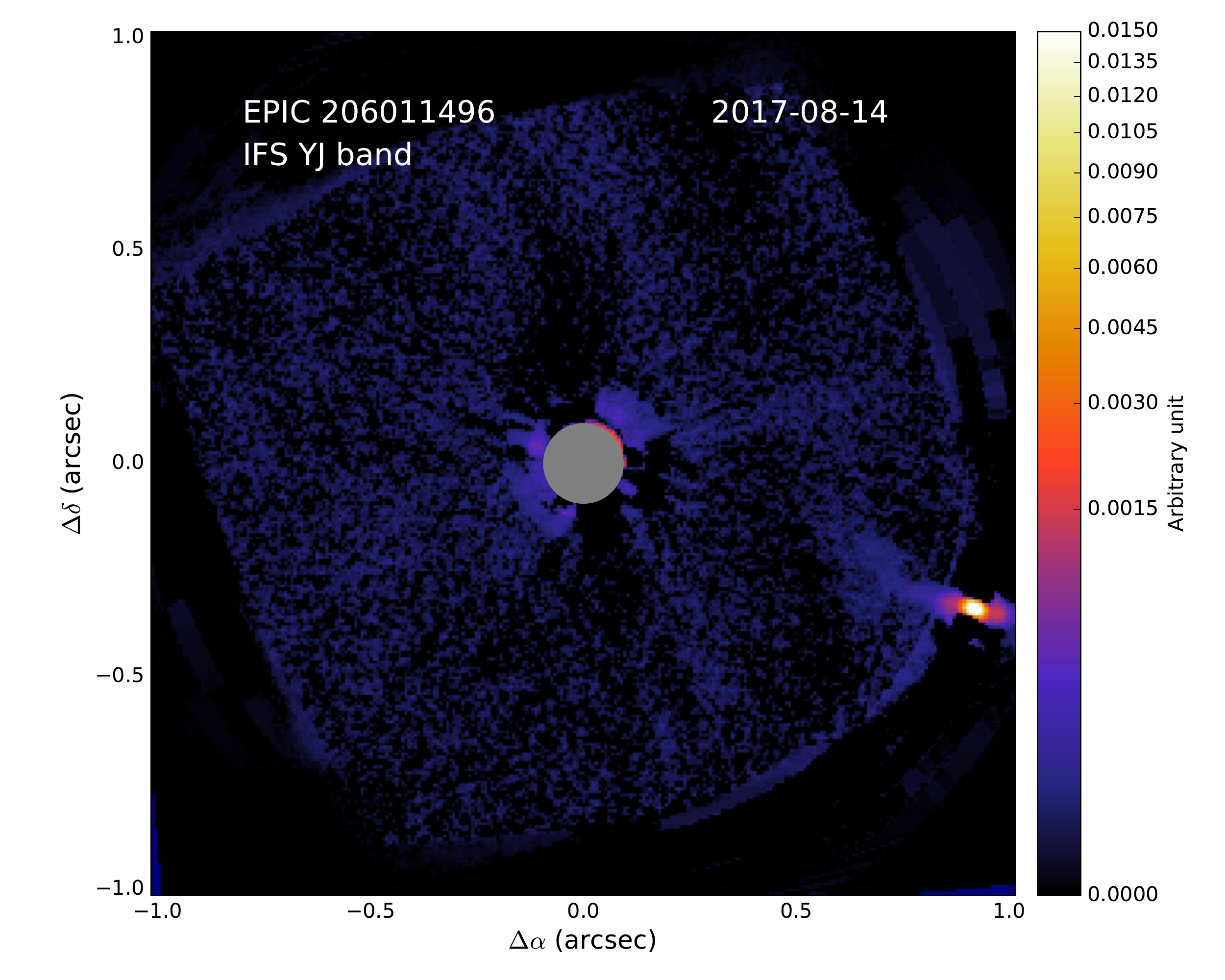} \\
\end{array}$
\end{center}
\hspace*{-1.30cm}
\caption{Combination of all-wavelength IFS images of the six \Kdeux targets, obtained with the TLOCI algorithm. The central gray disk represents the coronagraph. The companion EPIC\,206011496~B is well visible on the bottom right side of the image of EPIC\,206011496. North is up and east is left.}
\label{fig:allstars}
\end{figure*}

\subsection{No Detection around EPIC\,220383386, EPIC\,206157908, EPIC\,206144956, EPIC\,205904628 and EPIC\,206247743}
\label{sec:resultsall}

\begin{figure}
\centering
\includegraphics[width=8cm]{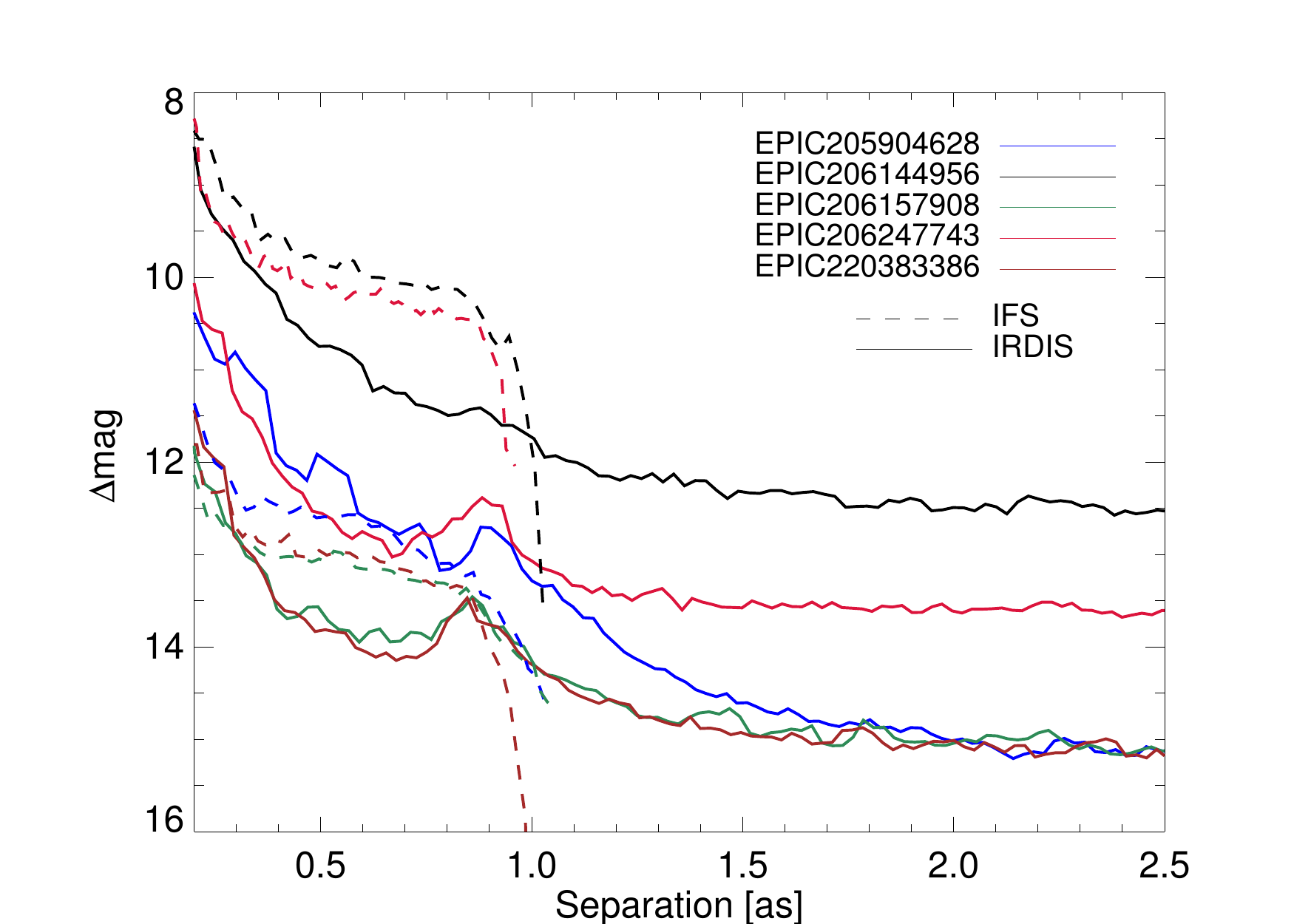}\\
\includegraphics[width=8cm]{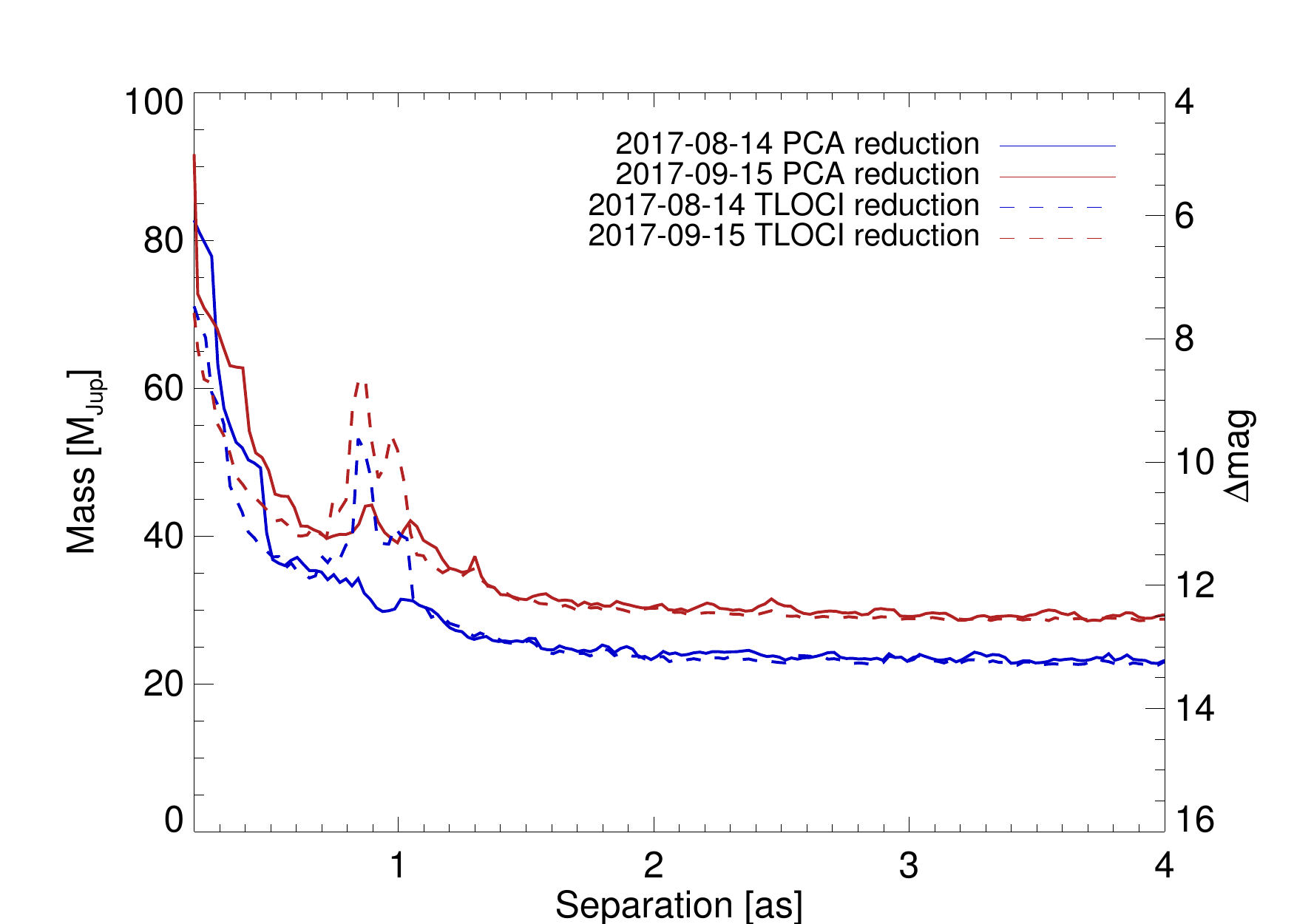}
\caption{Top: detection limits in $\Delta$mag for the five \Kdeux targets with no detection. The limits are calculated for IRDIS and IFS using the PCA algorithm. Bottom: detection limits for EPIC\,206011496 in mass and corresponding $\Delta$mag obtained with IRDIS.}
\label{fig:DetLim}
\end{figure}

Figure~\ref{fig:allstars} shows the IFS images of all six targets (IRDIS images can be found in Figure~\ref{fig:allstarsIRDIS} in Appendix \ref{app:imIRDIS}). We do not detect any candidate companion or any background star around five stars, but we detect a bright candidate around EPIC\,206011496 (see Section~\ref{sec:analyseEPIC206}). We calculated the detection limits for all five stars using the \textit{Specal} routine \citep{Galicher2018} offered by the SPHERE DC. Objects with contrasts between  $\sim$12.5 and 13.5 mag and separations between $0\farcs2$ and $6\arcsec$ in IRDIS data should have been detected (Figure~\ref{fig:DetLim}, top). In IFS data, for the same separation of $0\farcs2$, the detection limit is in the range of $\sim$8.5$-$12.5 mag. 

The detection limits are below the magnitudes of background eclipsing binaries that could mimic Earth-sized to Neptune-sized transits (in the case of a false-positive scenario). This means that our SPHERE observations entirely eliminate the possibility of such scenarios in the FoV covered by SPHERE, which drastically decreases the likelihood of false-positive scenarios, since the FoV covered by SPHERE is very large. Background binary or tertiary systems that cannot be detected with our observations because they are too faint could not have caused the transits by themselves. Had we detected background multiple systems, spending time on RV vetting on these \Kdeux candidates would have been useless. Only eclipsing binaries hidden behind the coronagraph cannot be detected, but this is very unlikely because the area covered by the coronagraph is tiny ($\sim36$ times smaller than the FoV). Our observations therefore significantly increase the chance that the detected transits are caused by real exoplanets, and encourage RV vetting to confirm them.

\subsection{Detection of an M Dwarf around EPIC\,206011496}
\label{sec:analyseEPIC206}

We detect a bright companion (EPIC\,206011496~B) close to EPIC\,206011496, both in IRDIS and IFS data for epoch 2017 August 14 (Figures~\ref{fig:allstars} and \ref{fig:allstarsIRDIS} in Appendix \ref{app:imIRDIS}) and in IRDIS data only for the two other epochs (Table~\ref{tab:obslog}, Appendix \ref{app:ObsLog}). We reach a median signal-to-noise (S/N) ratio of 66.5 in IFS ($YJ$ band) and average S/N of 84.6 in IRDIS ($H$ band) for this companion, for epoch 2017 August 14 (PCA reduction). We thus performed an astrometric and spectroscopic analysis of the companion in the following sections.

\subsubsection{Proper motion}
\label{sec:pm}

\begin{figure}
\centering
\includegraphics[width=8cm]{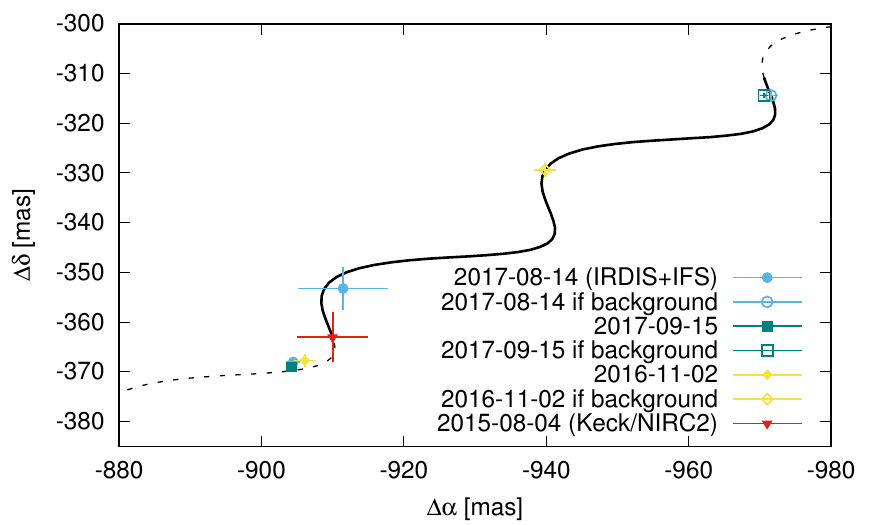}
\caption{Relative positions of the companion measured with VLT/SPHERE and Keck/NIRC2 with respect to the star EPIC\,206011496. The black line shows the apparent motion that a background star would have (solid: between observations; dashed: outside of them). The empty symbols show the theoretical positions of the companion at the observing periods if it was a background object. }
\label{fig:pm}
\end{figure}

We computed the position of the companion relative to the primary star between the datasets, using the parallax of the primary star and its proper motion (see Table~\ref{tab:stellarparam}), along with SPHERE astrometry from the three observing periods. We also added the position of the companion found with the NIRC2/Keck instrument (see Appendix~\ref{app:DetailsTargets}). The predicted positions of a hypothetical background object at the four observing periods are represented by empty symbols, while the measured positions are shown with plain symbols on Figure~\ref{fig:pm}. It is clear that the relative motion between the primary and the companion is insignificant. The companion is thus linked to the primary star and corresponds to the one detected with NIRC2/Keck in 2015. 

\subsubsection{Determination of EPIC\,206011496~A Parameters }
\label{sec:hoststarparam}

The parameters of EPIC\,206011496~B depend on the age of the system. We estimated the bolometric flux and the luminosity of the primary star using several photometric catalogs given by the VOSA tool \citep{Bayo2008}.\footnote{\url{http://svo2.cab.inta-csic.es/theory/vosa/}} In the case of EPIC\,206011496, we kept all photometric data points but the VISTA data, because they were flagged as bad data. We used the upper limit of the WISE.W4 data. The derived bolometric flux and associated luminosity are given in Table~\ref{tab:stellarparam}, and were obtained by combining the BT-NextGen AGSS2009 model \citep{Allard2011} and the \textit{Gaia} DR2 parallax. 
The best-fit model corresponds to an effective temperature of $T_{\rm eff}$  5400 $\pm$ 50~K, which is in very good agreement with the \textit{Gaia} DR2 temperature ($5390^{+194}_{-33}$~K) and within the error bars given by the model. However, it is lower than the previous determination of 5509 $\pm$ 50~K by \citet{Vanderburg2016b}. Their estimation was based on photometry and on the \textit{Hipparcos} distance (231~pc), which places EPIC\,206011496 much further away than \textit{Gaia} (139.22$_{-0.97 }^{+0.98}$ pc). They find a metallicity (0.07 $\pm$ 0.08 dex) compatible with ours, but both values are lower than the one derived by \cite{Huber2016}.
Using our derived $T_{\rm eff}$ and luminosity, we calculated the stellar radius to be $R = 0.92 \pm$ 0.02 $R_\odot$ using a standard propagation of errors. This value is much smaller than that of 1.714 $\pm$ 1.278 $R_\odot$ derived by \citet{Huber2016}, who also used \textit{Hipparcos} distance.\\
Finally, we used the PARSEC models \citep{Bressan2012} to derive the age and mass of the star. We used the technique described in \cite{Ligi2016} to interpolate the isochrones and compute the errors. 
As often, we obtained two different solutions: an old age of $2.42^{+3.76}_{-1.47} \times 10^9$ years with a 57$\%$ probability corresponding to a mass of 0.974 $\pm$ 0.044 $M_\odot$, and a younger age of $77.91^{+1.11}_{-0.46} \times 10^6$ years (43$\%$ prob.) with a mass of 0.995 $\pm$ 0.056 $M_\odot$ (Table~\ref{tab:stellarparam}). With no additional information on the star, we cannot choose between these two ages. 
Considering the derived probability and the lack of infrared excess in the spectral energy distribution (SED), we adopted the solution that corresponds to an evolved star of $2.42^{+3.76}_{-1.47}$~Gyr.

\begin{table}
	\centering
	\caption{Parameters of EPIC\,206011496}
	\label{tab:stellarparam}
	\begin{tabular}{lcl} 
	\hline \hline
	Parameter & Value & References \\
	\hline
	R.A. [hh mm ss]						 & 22 48 07.5629 & \textit{Gaia} DR1 (1) \\
	Decl. [deg mm ss]		 & $-$14 29 40.837 &  \textit{Gaia} DR1 (1)\\
	Parallax [mas]						 & 7.183 $\pm$ 0.051 & \textit{Gaia} DR2 (2) \\
	$\mu_{\alpha}$ [mas yr$^{-1}$]	 &	30.935	 $\pm$ 1.521 & \textit{Gaia} DR1	(1)\\
	$\mu_{\delta}$ [mas yr$^{-1}$]	 &	 $-$23.702 $\pm$ 0.971	&	\textit{Gaia} DR1	(1)\\
	$J$ [mag]								 & 9.726 $\pm$ 0.026 &  2MASS catalog (3)\\
	$H$ [mag]								 & 9.312 $\pm$ 0.022 &  2MASS catalog (3)\\ 
	$K$ [mag] 								 & 9.259 $\pm$ 0.027 & 2MASS catalog (3) \\
	$F_{\rm bol}$	[erg cm$^{-2}$ s$^{-1}$]  & 1.067 $\pm$ 0.0023 $\cdot$ 10$^{-9}$ & VOSA (4) \\
	$T_{\rm eff}$ [K]				 & 5400 $\pm$ 50		& VOSA (4) \\
	$L$ [$L_\odot$]						 & 0.646 $\pm$ 0.011 & VOSA  (4)  \\
	$[M/H]		$				&	0.0						& VOSA  (4)  \\
	$R$ [$R_\odot$]						 & 0.92 $\pm$ 0.02		& This work \\
	Age (old) [Gyr]				 &	 2.42$^{+3.76}_{-1.47}$ & This work \\
	Mass (old) [$M_\odot$]	 & 0.974 $\pm$ 0.044	&  This work\\
	Age (young) [Myr]					 &  $77.91^{+1.11}_{-0.46}$ &  This work\\
	Mass (young) [$M_\odot$]		 &	 0.995 $\pm$ 0.056	& This work\\
	\hline
	\end{tabular}
	\tablecomments{(1) \cite{GaiaCat2016} \,; (2) \cite{GaiaCat2018} (3) \,; \cite{CutriCat2003} \,; (4) \cite{Bayo2008}.}
\end{table}

\subsubsection{Spectral Analysis of EPIC\,206011496~B}

\begin{figure}
\centering
\includegraphics[width=7.5cm]{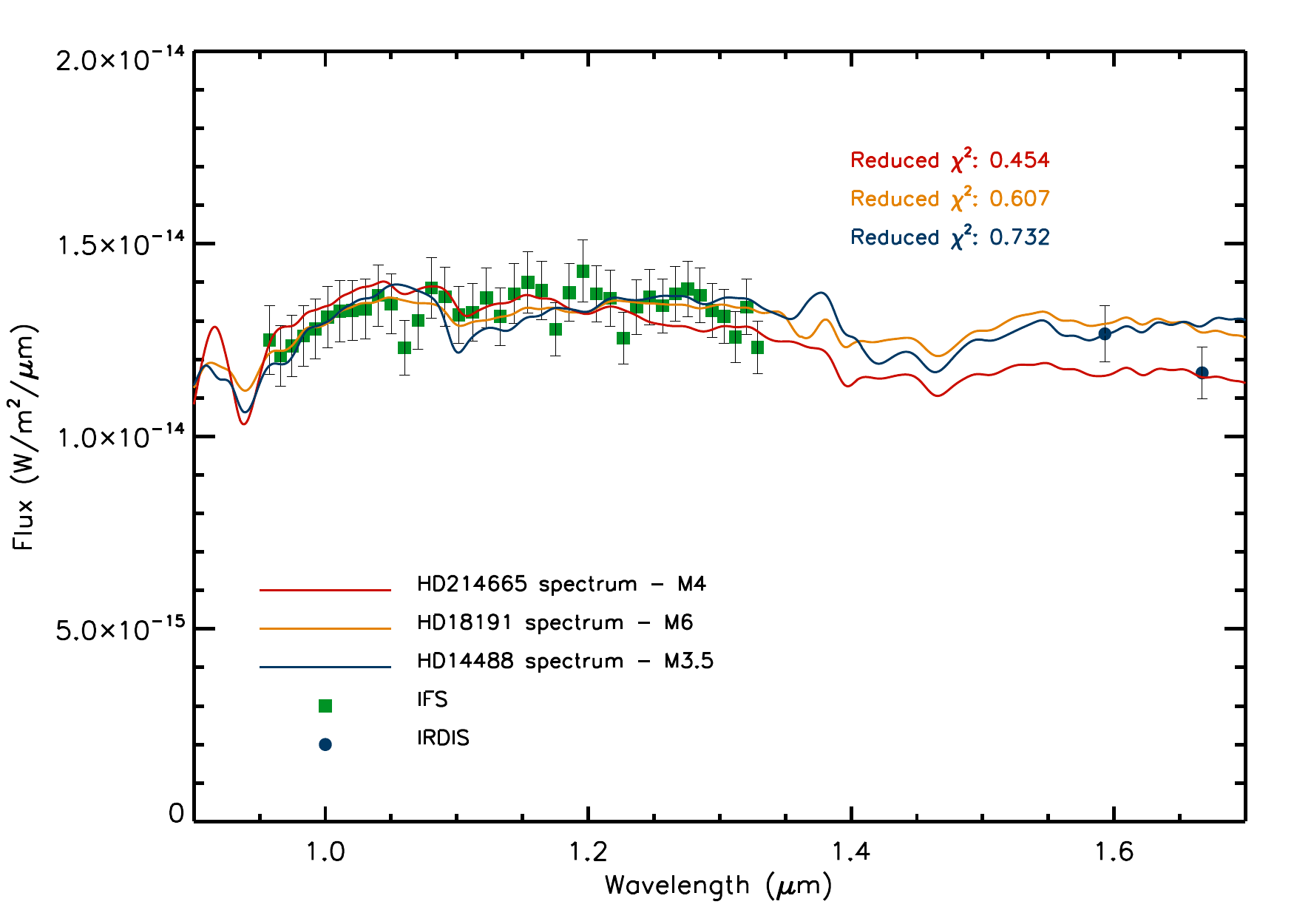}\\
\includegraphics[width=7.8cm]{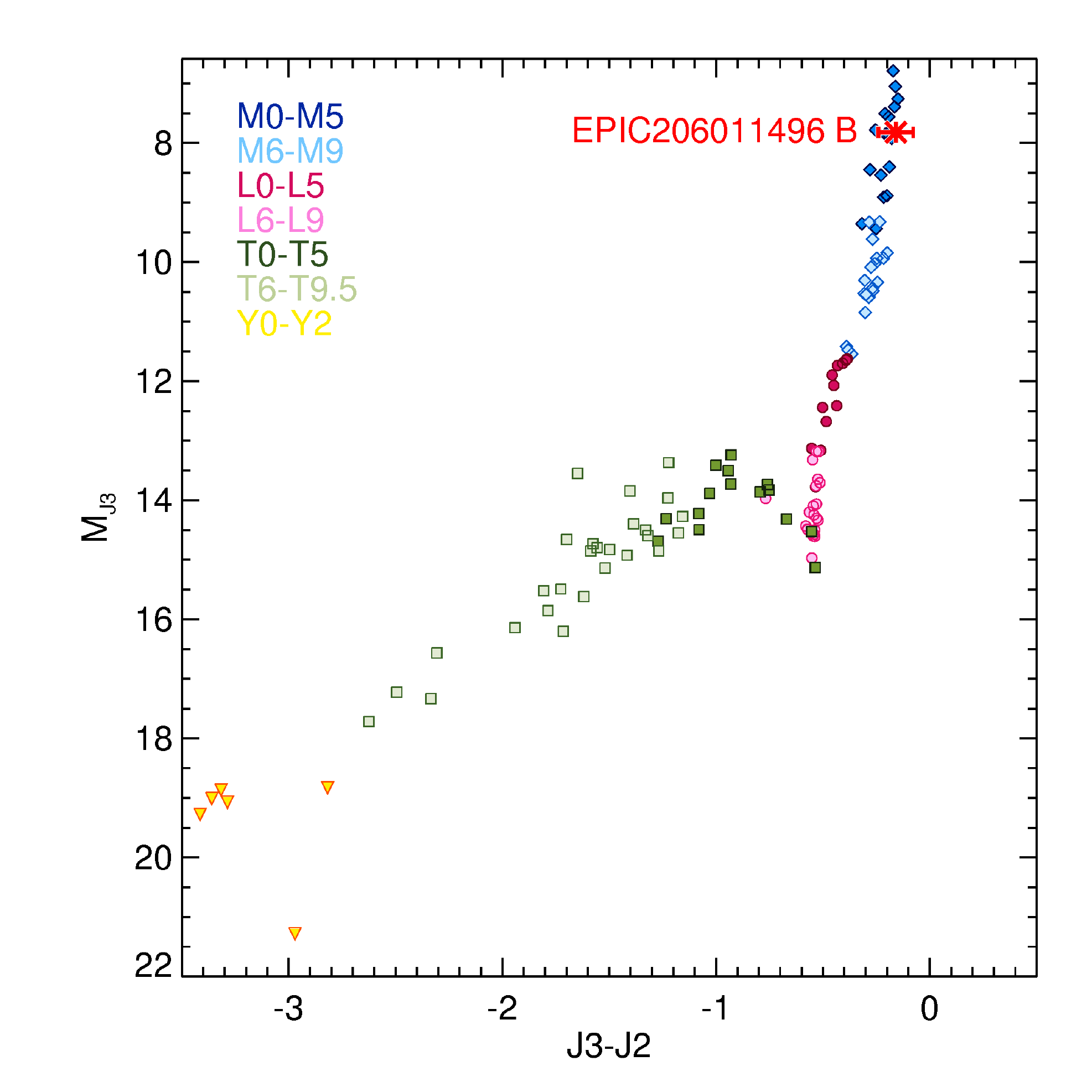}
\caption{Top: spectrum obtained with IFS and IRDIS (filled green and blue symbols) during period 2017 August 14 compared to the three best-fit M template spectra. Bottom: color-magnitude diagram using IFS photometry. EPIC\,206011496~B stands among the M0-M5 objects (red asterisk).}
\label{fig:spectra}
\end{figure}

Using both the IFS and IRDIS photometric values, we obtained a low-resolution
($R = 50$) spectrum of EPIC\,206011496~B in contrast. To convert it into flux, we first use a flux-calibrated BT-NEXTGEN \citep{Allard2012} synthetic spectrum of EPIC\,206011496~A, assuming $T_{\rm eff}$ = 5400~K, $\log(g)$ = 2.5, and [$M/H$] = 0.0 dex, which gives the best fit with the SED of EPIC\,206011496~A. We then multiply the flux in contrast by the synthetic spectrum of EPIC\,206011496~A.
The synthetic spectrum was also retrieved through the VOSA tool.
Since the system is probably old (see Section~\ref{sec:hoststarparam}), we made the fit to a library of M template spectra taken
from the library of \cite{Cushing2005} and \cite{Rayner2009}. The best fit is obtained for the 
M4 spectral type star HD\,214665 (see Figure~\ref{fig:spectra}, top, with $\chi^2 = 0.454$). We also tested the fitting with sample spectra of field dwarfs from the Spex Prism spectral Libraries \citep{Burgasser2014},\footnote{\url{http://pono.ucsd.edu/~adam/browndwarfs/spexprism/}} resulting in a best fit corresponding to the M7 star CTI021845.9+280047. With a fit done using the MLT field dwarfs from the IRTF library, we find a worse fit ($\chi^2 = 1.738$, GJ406 spectrum, M6 dwarf). As a result of this analysis, we adopt for EPIC\,206011496~B a spectral type of M4-7.

\subsubsection{Astrometry and Photometry}

\begin{table*}
\centering
\caption{Astrometry and Photometry of EPIC\,206011496 B}
\begin{tabular}{llllll}
\hline \hline
  	Filter		&			$\Delta$Mag     &   Mass   &    \multicolumn{2}{c}{Sep}    &       PA   \\
				&									&  ($M_{\odot}$)  &		(mas	)			& (au)					& 			($^\circ$)  \\	
				\hline
  $H2 $ &      3.19   $\pm$ 0.06    &   0.39   $\pm$   0.03    &    976.15   $\pm$     0.86    &    140.05    $\pm$   0.12    &    247.85   $\pm$   0.04 \\
  $H3$   &     3.13   $\pm$    0.06  &    0.38  $\pm$   0.03     &   976.75   $\pm$     0.97   &     140.13   $\pm$    0.14    &   247.87   $\pm$   0.04\\
  $Y$      &  4.39  $\pm$    0.02   &    0.30   $\pm$   0.03     &   977.90   $\pm$    0.27  &      140.30   $\pm$    0.04    &    248.81  $\pm$    0.01\\
  $J $     &  4.11  $\pm$    0.02     &  0.27   $\pm$   0.02     &   977.12   $\pm$     0.28    &    140.19   $\pm$    0.04      &  248.82  $\pm$    0.01\\

\hline
\multicolumn{6}{l}{Adopted Values (Medians and Standard Deviations)} \\
				 &    4.11 $\pm$ 0.64        &     0.38 $\pm$ 0.06       &     977.12 $\pm$ 0.73       &     140.19 $\pm$ 0.11         &    248.81 $\pm$ 0.55   \\						
\hline
\end{tabular}
\tablecomments{The contrast values in the $J2$ and $J3$ bands are 3.97 $\pm$ 0.05 and 3.81 $\pm$ 0.05, respectively. }
\label{tab:paramEPICB}
\end{table*}

The spectrum of the companion is also supported by its position in the color-magnitude diagram (CMD) built with field objects from the Spex Prism spectral Libraries \citep{Burgasser2014} and from \cite{Leggett2000}. A detailed description of how the CMD is built can be found in \cite{Bonnefoy2018}. In our analysis, the \textit{Gaia} DR2 parallaxes have been added for M and L field dwarfs. From the $H$ and $J$ photometry of the primary star, its distance, and the contrast of the companion, we computed its absolute magnitude in the $H$ and $J$ bands. In Figure~\ref{fig:spectra} (bottom), EPIC\,206011496~B is compared to field objects, and its position suggests an M0-M5 spectral type.

We computed the detection limit of the 2017 August 14 and 2017 September 15 observations of EPIC\,206011496 from IRDIS data in magnitude using the method described in \cite{Galicher2018}. Then, we used the COND 2003 model \citep{Baraffe2003} to convert the detection limits in mass, which depends on the age of the system, the distance, and the magnitude (Figure~\ref{fig:DetLim}). We only consider the old age. 
Using the COND model, we also derived the mass of the companion for each spectral band (Table~\ref{tab:paramEPICB}) using the method described in \cite{Bonavita2017}. We considered the median contrast in the wavelength range between  0.95 and 1.15 $\mu$m for the $Y$ band, and that between 1.15 and 1.35 $\mu$m for the $J$ band. We also estimated the companion separation and position angle for each band, with the uncertainties taking into account all the possible sources of errors. 
We finally derive a median mass of 0.38 $\pm$ 0.06 $M_{\odot}$ at a separation of 977.12 $\pm$ 0.73 mas for EPIC\,206011496~B from the data of epoch 2017 August 14. This mass estimation is compatible with the spectral type of M4-7, and is well above the detection limit in all our data.

\section{Summary and Discussion}
\label{sec:discussion}

We observed with SPHERE/VLT six \Kdeux targets that show transit light curves compatible with Earth-sized to Neptune-sized exoplanets. For EPIC\,205904628, EPIC\,206144956, EPIC\,206157908, EPIC\,206247743 and EPIC\,220383386 we do not detect any object in their close environments or in their backgrounds. 
With deep detection limits down to  $\sim$10 to 14 mag into \Kepler's PSF, the probability of such configurations as chanced-aligned eclipsing systems causing a TD in the light curves, is tremendously decreased.

We detect a companion around EPIC\,206011496 at a wide separation of 140.19 $\pm$ 0.11 au. Its spectrum is compatible with an M4-7 star and the proper motion analysis shows that it is bounded to the primary star. Given its separation, its orbital period (close to 7000 years) cannot cause the observed transit. Using COND evolutionary models, we estimate its mass to be 0.38 $\pm$ 0.06 $M_{\odot}$. Our SPHERE/VLT data are used to confirm, with Keck/NIRCS2 data, an exoplanetary candidate, which is presented in \cite{Lam2018}.

Using only our observations, we cannot unambiguously conclude whether the observed transit occurs on EPIC\,206011496~A or on EPIC\,206011496~B.
To compute the approximate transiting exoplanet radius, we consider the radius range (0.26$-$0.12 $R_{\odot}$) and temperature range (3100$-$2500~K) for M4-M7 dwarfs given by \cite{Kaltenegger2009}, and the average wavelength of the \Kdeux bandpass (660~nm). We then calculate the luminosity ratio of both stars at this wavelength using Planck's law. Assuming that the transit occurs on EPIC\,206011496~A and given the TD (Table~\ref{tab:transitinfo}, Appendix \ref{app:DetailsTargets}), the exoplanet would have a radius of $\sim1.59-1.62$ $R_{\oplus}$ considering the stellar radius in Table~\ref{tab:stellarparam} and the contamination by EPIC\,206011496~B. Here, the contamination by the companion is negligible as expected and the transiting planets remains in the super-Earth regime. If the exoplanet transited EPIC\,206011496~B, and taking the companion's radius between 0.12 and 0.26 $R_{\odot}$, the exoplanet radius would be included between $\sim2.17$ and $2.39 R_{\oplus}$. In this case, the exoplanets would fall into the mini-Neptune regime. In their paper, \cite{Lam2018} confirm that the planet transits the primary star.

As highlighted by \cite{Matson2018} and \cite{Kraus2016}, the impact of a stellar companion on planetary formation still remains an open question. Our SPHERE observations reveal a companion star in an exoplanetary system and thus contribute to the study of formation mechanisms, architecture, and binarity in exoplanetary systems, and could be integrated into larger imaging surveys \citep[like e.g.,][]{Bonavita2016, Matson2018}. We also provide deep imaging of the environment of \Kdeux targets, for which the vetting is still rare \citep{Matson2018}, and many exoplanetary candidates are not yet confirmed.

\acknowledgements

The authors thank the anonymous referee for his constructive comments.
R.L. has received funding from the European Union's Horizon 2020 research and innovation program under the Marie Sk\l odowska-Curie grant agreement n. 664931. R.L. thanks the Centre National d'\'Etudes Spatiales (CNES) for financial support through its post-doctoral program. 
O.D. acknowledges the support from Funda\c{c}\~ao para a Ci\^encia e Tecnologia (FCT) through national funds and by FEDER through COMPETE2020 by grants UID/FIS/04434/2013 \& POCI-01-0145-FEDER-007672 and PTDC/FIS-AST/1526/ 2014 \& POCI-01-0145-FEDER-016886. 
S.C.C.B. acknowledges support from FCT through national funds and by FEDER through COMPETE2020 and POCI by these grants UID/FIS/04434/2013 \& POCI-01-0145-FEDER-007672 \&  POCI-01-0145-FEDER-028953 and through Investigador FCT contract IF/01312/2014/CP1215/CT0004.
This work has made use of the SPHERE Data Centre, jointly operated by OSUG/IPAG (Grenoble), PYTHEAS/LAM/CeSAM (Marseille), OCA/Lagrange (Nice), and Observatoire de Paris/LESIA (Paris), and is supported by a grant from Labex OSUG@2020 (Investissements d'avenir - ANR10 LABX56). This work is based on observations collected at the European Organisation for Astronomical Research in the Southern Hemisphere under ESO programs 98.C-0779(A) and 99.C-0.279(A).
This publication makes use of VOSA, developed under the Spanish Virtual Observatory project supported from the Spanish MICINN through grant AyA2008-02156.
This work has made use of data from the European Space Agency (ESA)
mission {\it Gaia} (\url{https://www.cosmos.esa.int/gaia}), processed by
the {\it Gaia} Data Processing and Analysis Consortium (DPAC,
\url{https://www.cosmos.esa.int/web/gaia/dpac/consortium}). Funding
for the DPAC has been provided by national institutions, in particular
the institutions participating in the {\it Gaia} Multilateral Agreement.

\appendix

\section{Computation of the contrast of contaminants}
\label{app:calculContaminants}

The TD observed in a light curve, taking into account the presence of contaminants, is given by 
\begin{equation}
\delta_{\mathrm{obs}} = (1 - c) ~ \delta_{c=0} ,
\label{eq:transitdepth}
\end{equation}
where $\delta_{\mathrm{obs}}$ is the observed TD, $\delta_{c=0}$ is the TD in absence of contaminants, and $c$ is the contamination of the light curve. $c$ can be described as the percentage of the flux, in the photometric aperture, which comes from the contaminants (and thus not from the eclipsing system). It can be mathematically expressed as $c = F_{\mathrm{\textsc{c}}} / (F_{\mathrm{\textsc{c}}} + F_{\mathrm{\textsc{es}}})$, where $F_{\mathrm{\textsc{es}}}$ is the flux coming from the eclipsing system and $F_{\mathrm{\textsc{c}}}$ is the flux coming from all the contaminants. Equation~\ref{eq:transitdepth} can be transformed to obtain the difference in magnitude ($\Delta\mathrm{mag}$) between the flux from the eclipsing system and the flux from the contaminants necessary to produce an observed depth ($\delta_{\mathrm{obs}}$), assuming an uncontaminated TD ($\delta_{c=0}$):
\begin{equation}
\Delta\mathrm{mag} \simeq \mathrm{mag}(F_{\mathrm{\textsc{es}}}) - \mathrm{mag}(F_{\mathrm{\textsc{c}}} + F_{\mathrm{\textsc{es}}}) = - 2.5 \log(\delta_{\mathrm{obs}} / \delta_{c=0}) .
\end{equation}

We note that the higher the $\delta_{c=0}$, the higher the $\Delta\mathrm{mag}$. Consequently, the faintest eclipsing system that can mimic a planetary transit is 10 mag fainter than the target star for an Earth-sized transit ($\delta_{\mathrm{obs}} = 100$ ppm) and 7.5 mag fainter for a Neptune-sized transit ($\delta_{\mathrm{obs}} = 1000$ ppm).

\section{Details on the Sample of \Kdeux Targets}
\label{app:DetailsTargets}

In this section, we give more details on our \Kdeux targets, along with information on the transit detections (Table~\ref{tab:transitinfo}).

The two stars EPIC\,206144956 (BD-115779) and EPIC\,206011496 (BD-156276) show a V-shaped transit, typical of a grazing transit. For EPIC\,206144956, a planetary candidate of 1.65 \REarth was detected \citep{Vanderburg2016b}. 
Similarly, the light curve of EPIC\,206011496 reveals a transit signal corresponding to a 1.62 $\pm$ 0.12 \REarth exoplanet \citep{Vanderburg2016b}. \cite{crossfield197Candidates1042016} used \textsc{HRI} with the Keck/NIRC2 instrument as complementary observations, which did not allow them to validate the exoplanetary candidate. However, they mention another companion candidate of $\Delta\mathrm{mag}$ = 2.81 in the $K$ band at $0\farcs980$ separation. This candidate was also detected in other low-resolution Keck/NIRC2 observations from 2015 August 04 (program N151N2, PI Ciardi) at 2.169~$\mu$m. The astrometry provides a separation of 979 $\pm$ 5 mas and a PA of $248\fdg27 \pm 0\fdg29$, corresponding to $\Delta\alpha=-910\pm5$ mas and $\Delta\delta= -363\pm5$ mas \citep[see][for details]{Lam2018}. 

The system of EPIC\,220383386 (HD\,3167) is composed of three super-Earth-sized exoplanets. The first two, HD\,3167~b and HD\,3167~c, were recently discovered by \cite{Vanderburg2016a} with the transit method. They performed additional imaging follow-up using Robo-AO adaptive optics system \citep{Baranec2014, Law2014}. Their images allow contrasts of 2 mag at $0\farcs25$ from the star, and 5 mag at $1\arcsec$, and did not lead to any additional detection. However, using orbital analysis, they hypothesized that an additional non-transiting exoplanet could be part of the system. This was confirmed by RV measurements by \cite{Christiansen2017}, who also gave the densities of the two transiting exoplanets. \cite{christiansenThreeCompanyAdditional2017} performed additional \textsc{HRI} vetting with the Keck/NIRC2 camera, without the detection of an additional companion.

Concerning EPIC\,205904628 (HD\,212657) and EPIC\,206247743 (BD-096003), the TDs correspond to exoplanets of 2.13 \REarth and 1.67 \REarth, respectively. \cite{vaneylenK2ESPRINTProjectShortperiod2016} performed \textsc{HRI} follow-up observations of EPIC\,206247743 using the FastCam camera at the Telescopio Carlos Sánchez telscope and the Subaru telescopeʼs Infrared Camera and Spectrograph. Both observations led to no detection, but the reached contrasts did not allow us to discriminate between contaminants capable of mimicking a small planetary transit.

Finally, we have little information on EPIC\,206157908 (HD\,216252). The TD could correspond to an Earth-sized to Neptune-sized exoplanet, but no additional \textsc{HRI} follow-up observation has been performed to our knowledge.

\begin{table}
\centering
\caption{Targets and Results of the Light-curve Analysis}
\begin{tabular}{llllll}
\hline \hline
EPIC            & $Kp$    & \multicolumn{3}{c}{Transits Properties\tablenotemark{a} }\\ 
                    &          & $\mathrm{\delta}$ (ppm)  & $P$ (days)  & $T_{\rm 14}$ (h) &  References\\ 
\hline
205904628  & 8.2    & 275  &  9.9754      & 3.3	 & 1 \\
206011496  & 10.9  & 250  &  2.3684      & 2.5   & 2 \\
206144956  & 10.4  & 410  & 12.6530     & 3.5 	& 3 \\
206157908  & 9.4    & 700	  & 4.10	      &   ...	 & 4\\ 
206247743  & 10.6  & 432  & 4.6049       & 8.8	  & 5 \\
220383386  & 8.9    & 334	  & 1.0	      & 1.7	&  6\\
		   &          & 973  & 29.8 	      & 5.1 	& \\
\hline 
\end{tabular}
\tablecomments{
(1) Detected and announced in \citet{Vanderburg2016b} and validated by \citet{mayo275Candidates1492018}.
(2) Detected by our team and announced in \citet{barrosNewPlanetaryEclipsing2016}, \citet{Vanderburg2016b} and \citet[][including low contrast \textsc{HRI}]{crossfield197Candidates1042016}.
(3) Detected by our team and announced in \citet{barrosNewPlanetaryEclipsing2016} and \citet{Vanderburg2016b}, and validated by \citet{crossfield197Candidates1042016}. 
(4) Detected by D. Armstrong (2018, private communication).
(5) Detected and announced in \citet{Vanderburg2016b}, and confirmed as the planet K2-39 b by \citet[][including a low contrast \textsc{HRI}]{vaneylenK2ESPRINTProjectShortperiod2016} and \citet{petiguraFourSubSaturnsDissimilar2017}. Later observed by \citet{schmittPlanetHuntersSearching2016} with low contrast \textsc{HRI}.
(6) Detected, announced, and confirmed by \citet[][including low contrast \textsc{HRI}]{Vanderburg2016a} (i) and later by \citet{petiguraPlanetCandidatesK22017}, confirmed by \citet[][including low contrast \textsc{HRI}]{christiansenThreeCompanyAdditional2017}.
\tablenotetext{a}{$\mathrm{\delta}$ is the measured transit depth, $P$ is the period of the transits, and $T_{14}$ is the transit duration.}
}
\label{tab:transitinfo}
\end{table}

\section{Observing Journal}
\label{app:ObsLog}

\begin{table*}
	\centering
	\caption{Observing Log of SPHERE/VLT Observations}
	\label{tab:obslog}	
	\begin{tabular}{llcccccc} 
		\hline \hline
		Star & UT Date		& 		Instr.		&  	Filter		& 		 NDIT$\times$DIT	 & N frame\tablenotemark{b} & 	Field rotation	&	Seeing   \\
				&					&						&					&		  (s)										&				&	 ($^\circ$)  				& $\arcsec$\\
		\hline
		\multirow{2}{*}{EPIC\,220383386}  & \multirow{2}{*}{2016 Oct 08} 	 &		IFS	   	&	$YJ$ 	    &	1 $\times$ 64		& 33 &	23.18	&	0.64	\\
									& 						 &		IRDIS 	&	$H2H3$	&	3 $\times$ 64	&	 35  &	22.63	&	0.65	\\
		
		\multirow{2}{*}{EPIC\,206157908} & \multirow{2}{*}{2016 Oct 08}   & 	IFS	   &	$YJ$ 	    &	1 $\times$ 64	& 31  &	 44.18 	& 	0.70	 \\
									& 						 &		IRDIS 	&	$H2H3$	&	3 $\times$ 64	&	31 &	43.26	&	0.69	\\
		
	\multirow{2}{*}{EPIC\,206144956} & \multirow{2}{*}{2016 Oct 25} &		IFS	   &	$YJ$	        &	1 $\times$ 64	& 19 &	29.12	&	1.17	\\
									& 						&		IRDIS 	&	$H2H3$	&	3 $\times$ 64	       &	 21 &	28.22	       &	1.48	\\
		
	\multirow{2}{*}{EPIC\,206247743} & \multirow{2}{*}{2016 Oct 28} &		IFS	   &	$YJ$     	&	1 $\times$ 64	&	31  &	29.37	&	0.78	 \\
									& 						&		IRDIS 	&	$H2H3$	&	3 $\times$ 64	       &	 29 &		30.49       &	0.67	 \\
		
	\multirow{2}{*}{EPIC\,205904628} & \multirow{2}{*}{2017 Aug 29} &	IFS	   		&	$YJ$ 	   &	1 $\times$ 96	&	 23 &	108.41	&	0.81	\\
									& 						 &	 IRDIS 		&	$H2H3$	&	1 $\times$ 96	&	 21 &	109.60	&	0.73	 \\
		
		\multirow{2}{*}{EPIC\,206011496} & \multirow{2}{*}{2016 Nov 02} &	IFS	   		&	$YJ$			&	1 $\times$ 64	& 48	& 	44.56	&	0.84\\
									& 						 &	IRDIS 		&	$H2H3$	&	3 $\times$ 64		& 15 &	33.94	&	0.78	 \\
		
		
		\multirow{2}{*}{EPIC\,206011496} & \multirow{2}{*}{2017 Aug 14}&	IFS	   &	$YJ$			&	1 $\times$ 96	 & 9 &	14.49 &	0.76			\\
									& 						&	IRDIS 	&	$H2H3$	&	4 $\times$ 48 &	55 &	47.52 &	0.8	\\
		
		
		\multirow{2}{*}{EPIC\,206011496} & \multirow{2}{*}{2017 Sep 16} &	IFS	   &	$YJ$			&	1 $\times$ 96 & 21 &	 39.98	&	0.98	 \\
									& 						 &	 IRDIS 	&	$H2H3$	&	4 $\times$ 48	& 33 &	57.04	&	0.98 \\
\hline 
	\end{tabular}
	\tablecomments{The True North correction is equal to $-1\fdg75$. The plate scale for IRDIS is 12.27 and 7.46 mas pixel$^{-1}$ for IFS. }
	\tablenotetext{b}{After selection.}
\end{table*}

\section{IRDIS images}
\label{app:imIRDIS}

\begin{figure*}
\begin{center}$
\begin{array}{ccc}
\hspace*{-0.7cm}
\includegraphics[scale=0.25]{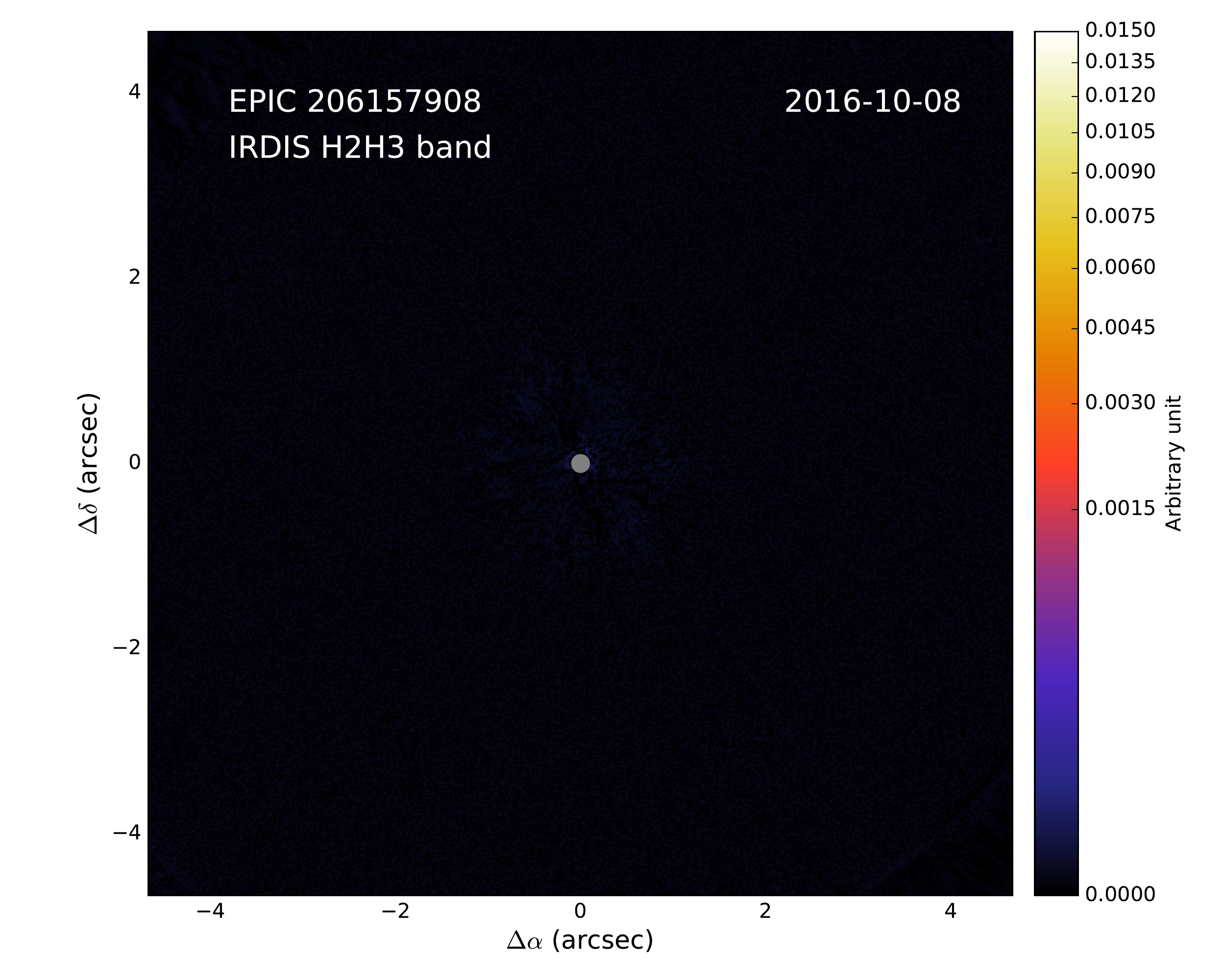} &
\hspace*{-0.5cm}
\includegraphics[scale=0.25]{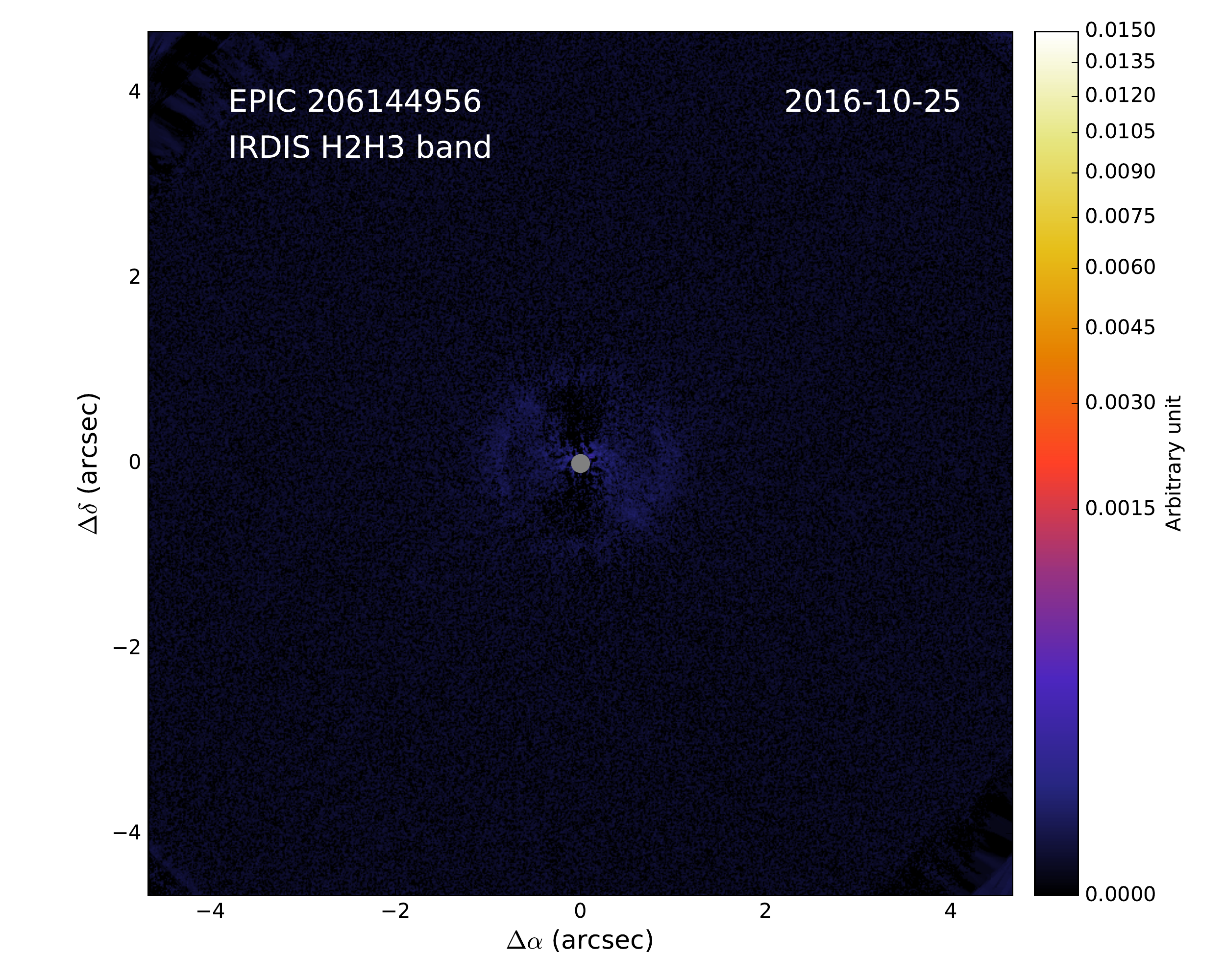} &
\hspace*{-0.5cm}
\includegraphics[scale=0.25]{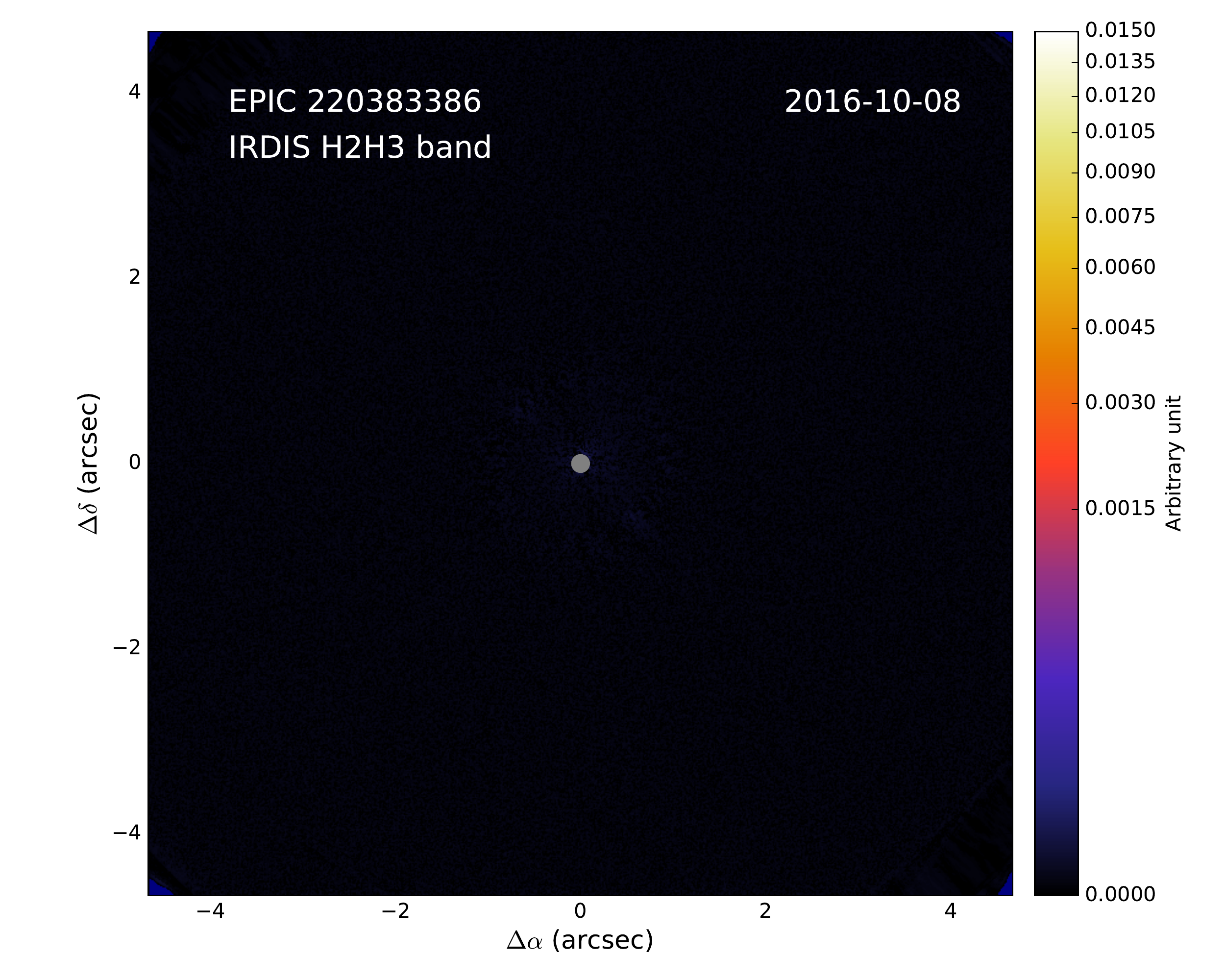} \\
\hspace*{-0.7cm}
\includegraphics[scale=0.25]{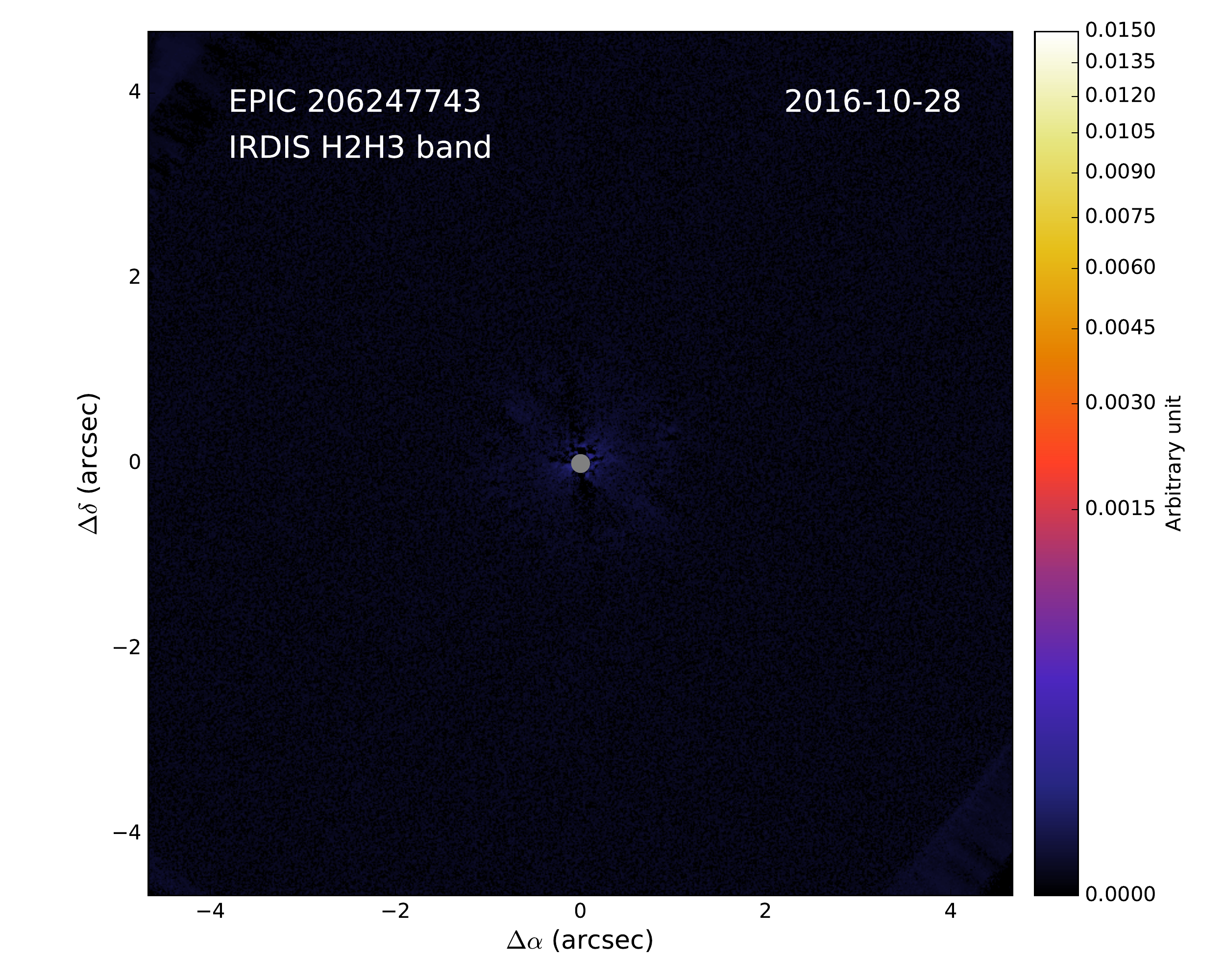} &
\hspace*{-0.5cm}
\includegraphics[scale=0.25]{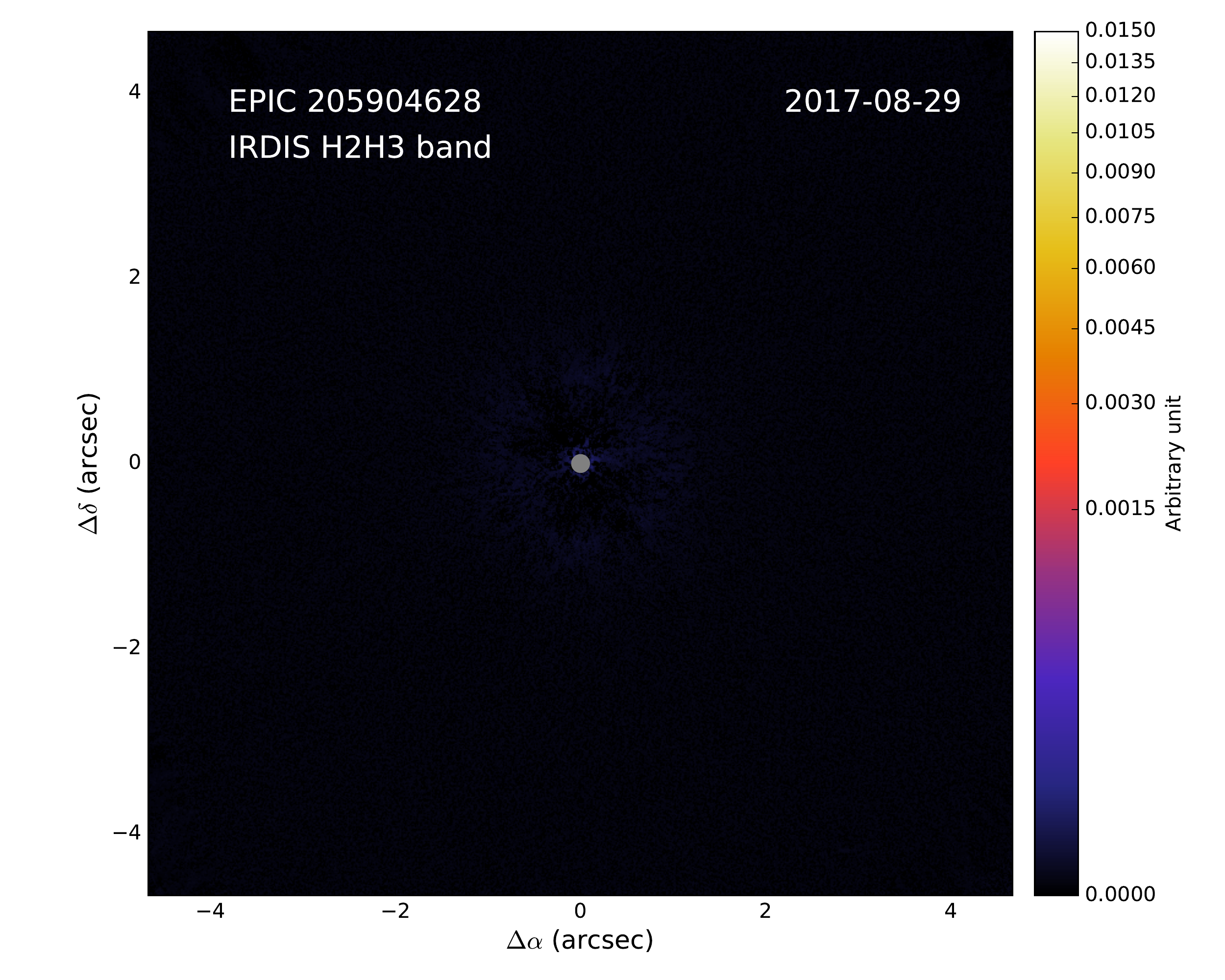} &
\hspace*{-0.5cm}
\includegraphics[scale=0.25]{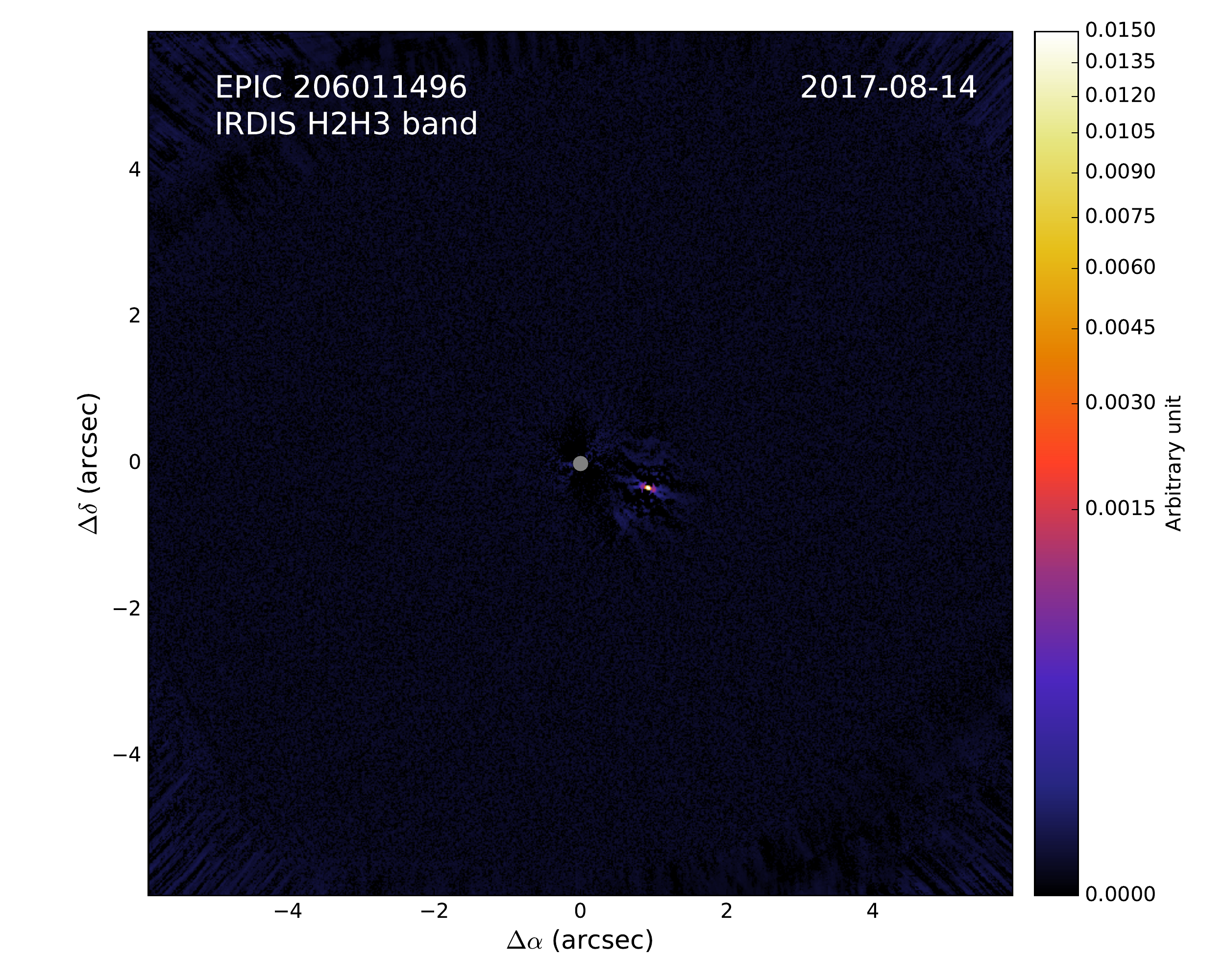} \\
\end{array}$
\end{center}
\hspace*{-1.30cm}
\caption{Combination of all-wavelength IRDIS images (slightly zoomed) of the six \Kdeux targets, obtained with the PCA algorithm (5 substracted modes). The central gray disk represents the coronagraph. The companion EPIC\,206011496~B is visible on the right side of EPIC\,206011496. North is up and east is left.}
\label{fig:allstarsIRDIS}
\end{figure*}

\bibliographystyle{aasjournal}
\bibliography{K2.bib}

\end{document}